\documentclass[11pt, fleqn]{article}

\usepackage[utf8]{inputenc}
\usepackage[T1]{fontenc}

\usepackage[a4paper, margin=2cm]{geometry}

\usepackage{mathtools,amssymb}
\usepackage{tensor}

\usepackage[dvipsnames]{xcolor}
\usepackage[colorlinks=true, allcolors=Blue]{hyperref}
\usepackage[capitalize]{cleveref}

\usepackage[hyphenbreaks]{breakurl}

\usepackage{lmodern}

\usepackage[sorting=none,url=true,backend=biber,style=numeric-comp,giveninits=true,doi=true,backref=false]{biblatex}

\bibliography{bibliography}

\usepackage{mathrsfs}
\def\mfdM{\mathscr{M}}
\def\mfdB{\mathscr{B}}
\def\metricL{\ell}
\def\bg{{(0)}}
\newcommand{\order}[1]{O(\epsilon^{#1})}

\def\LieD{\mathfrak L}
\let\t\tensor
\let\p\partial

\def\half{\tfrac{1}{2}}
\def\quarter{\tfrac{1}{4}}

\usepackage{authblk}

\title{\bfseries{Relativistic Theory of Elastic Bodies in the Presence of Gravitational Waves}}

\author[1]{Mario Hudelist}
\author[2]{Thomas B. Mieling}
\author[1]{Stefan Palenta}

\affil[1]{University of Vienna, Faculty of Physics, Boltzmanngasse~5, 1090 Vienna, Austria}
\affil[2]{University of Vienna, Faculty of Physics, Vienna Doctoral School in Physics (VDSP), Boltzmanngasse~5, 1090 Vienna, Austria}

\date{\today}

\usepackage[normalem]{ulem}
\newcommand{\add}[1]{{\color{black}#1}}

\begin{document}
\maketitle

\begin{abstract}
    The equations of motion governing small elastic oscillations of materials, induced by gravitational waves, are derived from the general framework of Carter and Quintana.
    In transverse-traceless gauge, no bulk forces are present, and the gravitational wave is found to act as an effective surface traction.
    For thin rods, an equivalent description is given, in which there is no surface traction, but a bulk acceleration, which is related to the Riemann curvature of the gravitational wave.
    The resulting equations are compared to those of the Synge--Bennoun elasticity theory.
\end{abstract}

\thispagestyle{empty}

\section{Introduction}

For the theoretical description of elastic materials in curved space-time, there are multiple approaches.
Here, we distinguish what we refer to as the \emph{Synge--Bennoun (SB) theory} from the \emph{Carter--Quintana (CQ) theory}.

The SB theory, essentially laid out in Refs.~\cite{Synge_1959,Bennoun_1964}, dispenses with a notion of \emph{strain} (which, in the classical theory, quantifies the deviation of the spatial metric from a reference configuration), but uses a notion of \emph{rate of strain}. Hooke’s law is then reinterpreted as not to provide a linear relationship between stress and strain, but rather as a linear relationship between \emph{rates} of stress and strain.

The CQ theory \cite{1972RSPSA.331...57C} on the other hand, introduces the notion of a three-dimensional \emph{material manifold}. By endowing this material manifold with a Riemannian metric, one is able to define a notion of strain, and hence can use direct generalisations of non-relativistic stress-strain relationships to obtain a stress tensor.
Even in the absence of a preferred Riemannian reference metric, this theory allows defining a notion of stress via what might be called a stress-deformation-relationship.
Moreover, unlike the SB theory, the CQ theory is not limited to the regime of linear elasticity, but also allows for strong deformations.

While most current texts on mathematical aspects of relativistic elasticity are based on the CQ theory (and are seemingly unaware of the SB theory), see e.g.\ Refs.~\cite{2003CQGra..20..889B,2004gr.qc.....3073B,2006Wernig-Pichler,2021CQGra..38h5017B} and references therein, articles focusing on experimental applications with regards to gravitational wave detection mostly rely on the SB theory instead \cite{AIHPA_1979__30_3_251_0,2007CRPhy...8...69V}.

The aim of this text is to describe the effect of weak gravitational waves on elastic objects using the CQ theory, and to compare the resulting description with the SB theory, and also with related equations put forward by various authors on heuristic grounds (in fact, Weber’s original papers were not based on any fully developed relativistic elasticity theory).

We start in \cref{s:setup} by laying out the mathematical description of elastic materials in a general curved space-time. In this setting, we naturally expect the material to be subject to deformations. These deformations determine the stresses, with details depending on the material properties. For the considered case of hyperelastic materials, this stress-deformation relationship is fully determined by an energy density function.
% (which coincides with Lagrangian for the theory up to a sign).
\Cref{s:perturbation} specialises the general theory to linearized elastic perturbations induced by weak, but otherwise arbitrary perturbations in the space-time metric.

\add{In \Cref{s:EOM for GW} we apply the general formalism developed in the previous sections to the case of plane gravitational waves, where we recover the classical Navier--Cauchy equations, supplemented by boundary conditions where the gravitational wave acts as an effective external traction force.}

To illustrate these equations, in \cref{s:thin rods} we explicitly solve the simple example of an elastic thin rod, responding to \add{gravitational} radiation.
Finally, we compare with Papapetrou’s equations for gravitational-wave-induced elastic oscillations in \cref{s:CQ vs SB}.

Compared to previous treatments of the problem within the CQ theory, this work emphasises the importance of boundary conditions at the material’s surface, without which no unique solution to the elasticity equations can be obtained.

This text uses geometric units in which the speed of light is set to unity, and the metric signature is taken to be “mostly positive.”

\section{Review of Relativistic Elasticity}
\label{s:setup}

This section reviews the theory development of general relativistic elasticity provided in Ref.~\cite{1992JGP.....9..207K}, while maintaining the definition of deformation and the notation of various deformation tensors from Ref.~\cite{2021CQGra..38h5017B}, which are closely related to the elasticity theory notions from classical continuum mechanics.
Other related works such as Refs.~\cite{2006Wernig-Pichler,2003CQGra..20..889B,BrodaStephan2008Cotd} follow essentially the same theory development, but use slightly altered nomenclatures and conventions.

\add{The main object in the CQ theory} of relativistic elasticity is the \emph{deformation map} $f: \mfdM \to \mfdB$, where $\mfdM$ is the four-dimensional space-time manifold (with local coordinates $\t x{^a}$) and $\mfdB$ is a three-dimensional manifold, the “body” (with local coordinates $\t X{^A}$). The configuration is required to be such that there is a unique unit timelike vector field $\t v{^a}$ (the material’s four-velocity) satisfying $\t\p{_a} \t f{^A} \t v{^a} = 0$.
Conceptionally, this means that $f$ assigns to each point in space-time the material point present there, and the world-lines of material points are given by lines of constant $f$.

The velocity vector field gives rise to the spatial projection $\t*\metricL{^a_b}$ and the spatial metric $\t\metricL{_a_b}$ via
\begin{align}
	\t*\metricL{^a_b} &= \t*\delta{^a_b} + \t v{^a} \t v{_b}\,,
	&
	\t\metricL{_a_b} &= \t g{_a_b} + \t v{_a} \t v{_b}\,.
\end{align}
where $g_{ab}$ is the space-time metric. The tensor field $\t*\metricL{^a_b}$ projects vectors onto $v_\perp$, the subspace orthogonal to $\t v{^a}$, and $\t\metricL{_a_b}$ is the restriction of the space-time metric to $v_\perp$ and thus encodes the spatial geometry of the medium.
One then defines the spatial deformation gradient $\t H{^A_a}$ and the material deformation gradient $\t F{^a_A}$ through the equations
\begin{align}
    \label{eq:deformation gradients}
	\t H{^A_a} &= \t\p{_a} \t f{^A}\,,
	&
	\t H{^A_c} \t F{^c_B} &= \t*\delta{^A_B}\,,
	&
	\t F{^a_C} \t H{^C_b} &= \t*\metricL{^a_b}\,.
\end{align}
These quantities can be viewed as mappings which translate tensor fields between the configuration manifold $\mfdB$ and the space-time $\mfdM$. Starting from the deformation gradient, one can introduce various types of deformation tensors as in classical continuum mechanics.

For the purposes of this paper, we will restrict the discussion to the Green deformation $\t C{_A_B}$ and the Piola deformation $\t B{^A^B}$, defined as
\begin{align}
    \label{eq:deformation tensors def}
	\t C{_A_B} &= \t F{^a_A} \t F{^b_B} \t\metricL{_a_b}\,,
	&
	\t B{^A^B} &= \t H{^A_a} \t H{^B_b} \t\metricL{^a^b}\,,
\end{align}
which also encode the spatial geometry and are essential to obtain a notion of strain. Since $\t*\metricL{_a_b}$ and $\t*\metricL{^a^b}$ are inverse to each other, \cref{eq:deformation gradients} implies that $\t C{_A_B}$ and $\t B{^A^B}$ are mutually inverse.

To further quantify compression of elastic bodies, and to formulate a continuity equation, a volume-form on $\mfdB$ is necessary.
If $\Omega$ is such a form, then its pull-back along the configuration, $f^* \Omega$, is necessarily proportional to the spatial volume form $\phi = v \lrcorner\, \varPhi$, where $\varPhi$ is the space-time volume form and $\lrcorner$ denotes the interior product. The Jacobian $J$ is then defined as the proportionality factor in
\begin{equation}
    \phi = J f^* \Omega\,.
\end{equation}
Here, we take $\Omega$ to have units of volume, so that $J$ is dimensionless.
The Jacobian $J$ thus compares the “actual volume” $\phi$ with the “reference volume” $\Omega$ and thus measures expansion ($J > 1$) or contraction ($J < 1$).
As $\Omega$ is closed, it follows that $J^{-1} \phi$ is closed as well, which is equivalent to the continuity equation
\begin{equation}
    \t\nabla{_a}(J^{-1} \t v{^a}) = 0\,.
\end{equation}

Further, for hyperelastic media, the energy density $\rho$ (as measured in the rest-frame of the material) has the form
\begin{equation}
    \label{eq:energy density mass and elastic energy}
    \rho = J^{-1}(\varrho + W )\,,
\end{equation}
where $\varrho = \varrho(\t X{^A})$ is the rest-mass energy density, and $W$ is the elastic energy density, which is required to depend on the space-time metric only via the Green deformation $\t C{_A_B}$, i.e.\ $W = W(\t X{^A}, \t C{_A_B})$.
This requirement corresponds to the principle of material frame indifference in classical elasticity \cite[Sect.~3.1.2]{2006Wernig-Pichler}.

In the CQ theory, the stress-deformation relationship is provided by the Doyle--Ericksen formula, which expresses the stress in terms of derivatives of the elastic energy with respect to the deformation. Specifically, one obtains the following expressions for the second Piola--Kirchhoff stress $\t S{^A^B}$ and the associated Cauchy stress $\t\sigma{^a^b}$:
\begin{align}
    \label{eq:Doyle-Erickson S Sigma}
     \t S{^A^B}
        &= 2 \frac{\p W}{\p \t C{_A_B}}\,,
    &
    \t\sigma{^a^b}
        &= J^{-1} \t F{^a_A} \t F{^b_B} \t S{^A^B}\,.
\end{align}
The equations of relativistic hyperelasticity are then given as the Euler--Lagrange equations associated to the Lagrangian $\mathscr L = - \rho$.
The corresponding Hilbert stress tensor then evaluates to
\begin{equation}
\label{HilbertStress}
    \t T{^a^b}
        \equiv 2 \frac{\p \mathscr L}{\p \t g{_a_b}} + \mathscr L \t g{^a^b}
        = \rho\, \t v{^a} \t v{^b} - \t \sigma{^a^b}\,,
\end{equation}
and one can show that the Euler--Lagrange equations for $\mathscr L = - \rho$ are equivalent to $\t T{^a^b}$ being divergence-free \cite{1992JGP.....9..207K}.

The bulk equations of motion, $\t\nabla{_a} \t T{^a^b} = 0$ must be supplemented by boundary conditions at the material’s surface.
The “natural” boundary conditions for the variational problem based on the Lagrangian $\mathscr L = - \rho$ are those of free boundaries \cite[Sect.~5]{2021CQGra..38h5017B}, which means that
\begin{equation}
    \label{eq:free boundary}
	\t T{^a^b} \t n{_b} |_{\p U} = 0\,,
\end{equation}
where $U$ is the space-time volume occupied by the body, and \add{$\t n{_b}$} is the unit conormal to its boundary.

\begin{figure}
    \centering
    \includegraphics[width=0.7\textwidth]{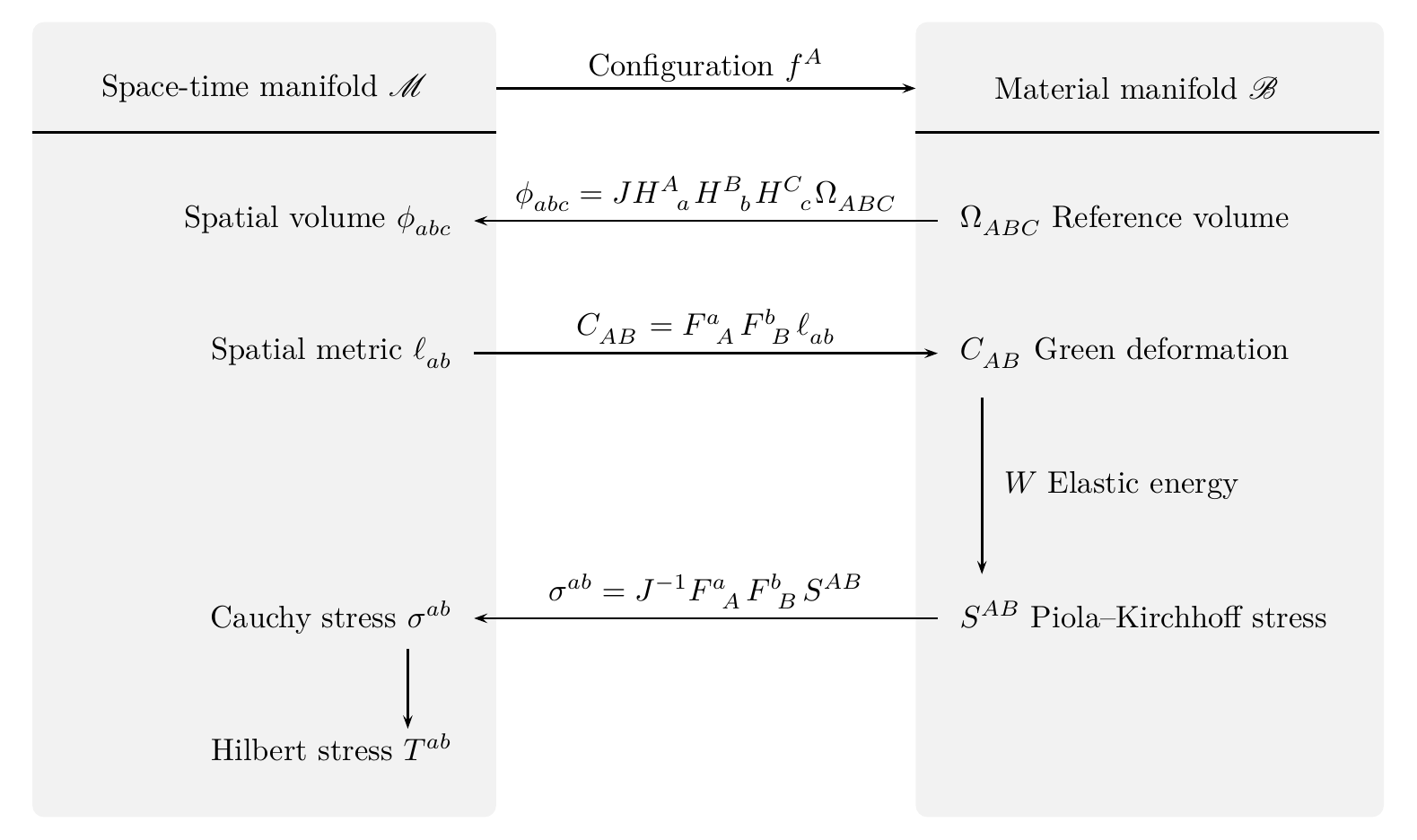}
    \caption{Schematic overview of the mathematical structure of elasticity theory.}
    \label{fig:my_label}
\end{figure}

We summarise the mathematical structure graphically in \cref{fig:my_label}.
The configuration of an elastic material is described by a mapping $f$ from the space-time $\mfdM$ to the material manifold $\mfdB$ (with certain properties). The differential of $f$ determines the deformation gradients $\t H{^A_a}$ and $\t F{^a_A}$, which allow to transport geometric objects between the two manifolds $\mfdM$ and $\mfdB$.
In particular, the spatial metric $\t\ell{_a_b}$ determines the Green deformation $\t C{_A_B}$ on $\mfdB$, which, by the Dolye--Ericksen formula, gives rise to the second Piola--Kirchhoff stress $\t S{^A^B}$. The corresponding quantity on the space-time manifold $\mfdM$ is the Cauchy stress $\t\sigma{^a^b}$, which determines the Hilbert stress via $\t T{^a^b} = \rho \t v{^a} \t v{^b} - \t \sigma{^a^b}$, where $\t v{^a}$ is the material’s four-velocity and $\rho$ is its energy density.
This shows that the stress-deformation relationship is implemented as a relation between tensors on $\mfdB$, which are translated to tensors on $\mfdM$ using deformation gradients.

\section{Elastic Deformations Induced by Metric Perturbations}
\label{s:perturbation}

This section develops the general \add{equations} necessary to describe elastic perturbations induced by small deviations of the space-time metric from the flat Minkowski metric.
Specifically, let $\epsilon$ be a small parameter characterising the metric perturbation and consider a metric of the form
\begin{equation}
    \t g{_a_b}
        = \t\eta{_a_b}
        + \t h{_a_b}
        + \order{2}\,,
\end{equation}
where $\t\eta{_a_b}$ is the Minkowski metric and $\t h{_a_b}$ is a small perturbation of order $\epsilon$.

To describe elastic perturbations induced by the metric perturbation $\t h{_a_b}$, starting from a configuration describing an inertial material in flat space-time, consider configurations of the form
\begin{equation}
    \label{configpert}
    \t f{^A}(\t x{^a})
        = \t*f{_\bg^A}(\t x{^a})
        - \t u{^A}(\t x{^a})
        + \order{2}\,.
\end{equation}
where $(\t*f{_\bg^A}(t,x,y,z)) = (x, y, z)$, which we write as \add{$\t* f{_\bg^A} = \t x{^A} \equiv \t*\delta{^A_a} \t x{^a}$}, describes an inertial material in flat space-time and $\t u{^A}$ is a displacement of first order in $\epsilon$.

The associated configuration gradient $\t H{^A_a}$ and four-velocity $\t v{^a}$ take the form 
\begin{align}
    \t H{^A_a}
        &= \t*\delta{^A_a} - \t \p{_a}\t u{^A}
        + \order{2}\,,
    &
    (\t v{^a})
        &= (1 + \half \t h{_0_0}, \t{\dot u}{^A})
        + \order{2}\,,
\end{align}
where overset dots indicate partial time derivatives.

The unperturbed Green deformation reduces to the flat Euclidean metric $\t* C{^\bg_A_B} = \t\delta{_A_B}$, which can be regarded as a reference metric on $\mfdB$.
Via the configuration gradient, this corresponds to the Cauchy deformation
\begin{equation}
	\t c{_a_b}
		= \t H{^A_a} \t H{^B_b} \t*\delta{_A_B}\,.
\end{equation}
One can then define the Green strain $\t E{_A_B}$ and the Almansi strain $\t e{_a_b}$ as measures of deviation of the spatial metric from the reference geometry
\begin{align}
	\t E{_A_B} &= \half(\t C{_A_B} - \t\delta{_A_B})\,,
	&
	\t e{_a_b} &= \half(\t\metricL{_a_b} - \t c{_a_b})\,.
\end{align}
These two strain tensors are related via $\t E{_A_B} = \t F{^a_A} \t F{^b_B} \t e{_a_b}$ and $\t e{_a_b} = \t H{^A_a} \t H{^B_b} \t E{_A_B}$.
To first order, only the spatial components of $\t e{_a_b}$ are non-zero with
\begin{equation}
    \label{eq:strain perturbation}
    \t e{_i_j}
        = \half(
            \t\p{_i} \t u{_j}
            + \t\p{_j} \t u{_i}
            + \t h{_i_j}
        ) + \order{2}\,,
\end{equation}
which differs from the analogous expression in classical elasticity by the presence of the metric perturbation.

To compute the stress tensor $\t \sigma{^i^j}$, one may proceed as follows.
Using $\t\ell{^i^j} = \t\delta{^i^j} - \t h{^i^j} + O(\epsilon^2)$, \cref{eq:deformation tensors def} yields
\begin{equation}
    \t B{^A^B}
        = \t\delta{^A^B}
        - \t\p{^A} \t u{^B}
        - \t\p{^B} \t u{^A}
        - \t h{^A^B}
        + O(\epsilon^2)\,,
\end{equation}
where $\t\p{_A} = \t*\delta{_A^a} \t\p{_a}$ and indices are raised and lowered using the unperturbed metric $\t\delta{_A_B}$.
As the Green deformation $\t C{_A_B}$ is inverse to the Piola deformation $\t B{^A^B}$, one finds
\begin{equation}
    \t C{_A_B}
        = \t\delta{_A_B}
        + \t\p{_A} \t u{_B}
        + \t\p{_B} \t u{_A}
        + \t h{_A_B}
      + O(\epsilon^2)\,.
\end{equation}
Next, expanding \cref{eq:Doyle-Erickson S Sigma}, one obtains
\begin{equation}
    \label{eq:Piola stress expansion}
    \begin{split}
        \t S{^A^B}
            &= 2 \frac{\p W}{\p \t C{_A_B}}\bigg|_{\t C{_E_F} = \t\delta{_E_F} + \t\p{_E}\t u{_F} + \t\p{_F}\t u{_E} + \t h{_E_F} + O(\epsilon^2)}
            \\
            &= 2 \frac{\p W}{\p \t C{_A_B}}\bigg|_{\t C{_E_F} = \t\delta{_E_F}}
            + 2 \frac{\p^2 W}{\p \t C{_A_B} \p \t C{_C_D}}\bigg|_{\t C{_E_F} = \t\delta{_E_F}}[\t\p{_C}\t u{_D} + \t\p{_D}\t u{_C} + \t h{_C_D}]
            + O(\epsilon^2)\,,
    \end{split}
\end{equation}
so that it suffices to specify $W$ up to its second derivative with respect to $\t C{_A_B}$ at the reference configuration $\t \delta{_A_B}$.
Here, we make the following assumptions.
\begin{itemize}
    \item In the absence of deformations, the elastic energy density $W$ vanishes: $W|_{C = \delta} = 0$. (This assumption is not essential: any non-zero value of $W|_{C = \delta}$ can be absorbed in the rest-mass energy density $\varrho$, see \cref{eq:energy density mass and elastic energy}).
    \item The unperturbed configuration is stress-free, so that the first order term in \cref{eq:Piola stress expansion} vanishes. This is equivalent to $\t S{^A^B}|_{C = \delta} \equiv 2 \frac{\p W}{\p \t C{_A_B}}|_{C = \delta} = 0$.
    \item Finally, we assume the stiffness tensor $\t C{^A^B^C^D} = 4 \frac{\p^2 W}{\p \t C{_A_B} \p \t C{_C_D}} \big|_{C = \delta}$ to be that of a homogeneous and isotropic material, i.e.\ 
    \begin{equation}
        \t C{^A^B^C^D}
            = \lambda\, \t\delta{^A^B} \t\delta{^C^D}
            + \mu(
                \t\delta{^A^C} \t\delta{^B^D}
                + \t\delta{^A^D} \t\delta{^B^C}
            )\,,
    \end{equation}
    where $\lambda$ and $\mu$ are the material’s Lamé parameters \cite[Eq.~(6.2.23)]{EringenAhmedCemal1967Moc}.
\end{itemize}
Using \cref{eq:Doyle-Erickson S Sigma}, the explicit form of $\t e{_i_j}$, and $J = 1 + \order{}$, one finds the Cauchy stress to be given by
\begin{equation}
    \label{eq:stress from strain}
    \t \sigma{^i^j}
        = \lambda\, \t\delta{^i^j} \t e{^k_k}
            + 2 \mu \t e{^i^j}
            + \order{2}\,,
\end{equation}
which resembles Hooke’s law from classical elasticity, but is not identical to it, as the strain $\t e{_i_j}$ contains contributions from the metric perturbation.

With this at hand, we can now derive the bulk equations of motions.
Using \cref{HilbertStress} and ${\t\nabla{_a} (\rho \t v{^a}) = 0}$, one finds
\add{\begin{equation}
    \t\nabla{_a} \t T{^a^b}
        = \rho \t v{^a} \t\nabla{_a} \t v{^b}
        - \t\nabla{_a} \t \sigma{^a^b}
        = \rho (\t v{^a} \t\p{_a} \t v{^b} + \t\Gamma{^b_a_c} \t v{^a} \t v{^c})
        - (\t\p{_a} \t\sigma{^a^b} + \t\Gamma{^a_a_c} \t\sigma{^c^b} + \t\Gamma{^b_a_c} \t\sigma{^a^c} )\,.
\end{equation}}

Since $\rho = \varrho + O(\epsilon)$, $(\t v{^a}) = (1, 0) + O(\epsilon)$, $\t\p{_a} \t v{^b} = O(\epsilon)$ , $\t\Gamma{^a_b_c} = O(\epsilon)$ and $\t\sigma{^a^b} = O(\epsilon)$, this reduces to
\begin{equation}
    \label{eq:EOM intermediate}
    \t\nabla{_a} \t T{^a^b}
        = \varrho ( \t\p{_t} \t v{^b} + \t\Gamma{^b_0_0} ) - \t\p{_a} \t\sigma{^a^b} + O(\epsilon^2)\,.
\end{equation}
The equation $\t\nabla{_a} \t T{^a^0} = 0$ is identically satisfied since $\t v{^0} = 1 + \tfrac{1}{2} \t h{_0_0} + O(\epsilon^2)$, $\t\Gamma{^0_0_0} = - \tfrac{1}{2} \t\p{_t} \t h{_0_0} + O(\epsilon^2)$ and $\t \sigma{^0^a} = O(\epsilon^{2})$. This is in agreement with the general result that \add{$\t v{_a} \t\nabla{_b} \t T{^a^b} = 0$} is identically satisfied by construction of $\t T{^a^b}$ \cite[Sect.~3.1.4]{2006Wernig-Pichler}.
Finally, the equations $\t\nabla{_a} \t T{^a^i} = 0$ yield
\begin{equation}
    \label{eq:EOM general}
    \varrho (\t{\ddot u}{^i} + \t\Gamma{^i_0_0})
        - \t\p{_j} \t\sigma{^i^j}
        = \order{2}\,,
\end{equation}
where, as above, $\varrho$ is the (unperturbed) rest-mass density, and $\t \Gamma{^i_0_0} = \t\p{_0} \t h{^i_0} - \half \t\p{^i} \t h{_0_0}$.
The metric perturbation $\t h{_a_b}$ thus enters the equations of motion in two ways: once via the Christoffel symbols coming from the acceleration $\t a{^b} = \t v{^a} \t\nabla{_a} \t v{^b}$ and once via the strain tensor $\t e{_i_j}$ given in \cref{eq:strain perturbation}.

\section{Gravitational Waves}
\label{s:EOM for GW}

The general equations of the previous section can now be specialised to the particular case of gravitational waves in transverse-traceless gauge (TT gauge), where
\begin{align}
	\square \t h{_a_b} &= 0\,,
	&
	\t h{_0_a} &= 0\,,
	&
	\t\p{_a} \t h{^a_b} &= 0\,,
	&
	\t h{^a_a} &= 0\,.
\end{align}
The equations of motion given in \cref{eq:EOM general} now reduce to 
\begin{equation}
	\varrho \t{\ddot u}{^i} - \t\p{_j} \t\sigma{^j^i}
	= \order{2}\,.
\end{equation}
Since the divergence of the stress tensor is expressible using only the divergence of the strain and the gradient of the trace of the strain, the transverse and traceless metric perturbation drops out of the field equations.
Thus, in TT gauge, the equations of motion are identical to the standard Navier--Cauchy equations
\begin{equation}
	\varrho \t{\ddot u}{^i}
		= \mu \Delta \t u{^i}
		+ (\lambda + \mu) \t\p{^i} \t\p{_j} \t u{^j}\,,
\end{equation}
in which the gravitational wave does not \add{enter}.

However, it would be wrong to conclude that gravitational waves have no effect on elastic bodies, when described in TT coordinates.
Indeed, the bulk equations of motion must be supplemented by boundary conditions to form a complete set of equations.

In the absence of external forces, \cref{eq:free boundary} applies, and yields
\begin{equation}\label{bc}
	\t \sigma{^i^j} \t n{_j} |_{\p U} = \order{2}\,,
\end{equation}
where \add{$\t n{_j}$} is the \add{spatial} unit conormal to the material’s boundary $\p U$.
In this equation, the gravitational wave contribution does not cancel.

To interpret this result, it is convenient to define the “classical strain” $\t{\underline e}{_i_j}$ and “classical stress” $\t{\underline \sigma}{^i^j}$ as
\begin{align}
	\t{\underline e}{_i_j}
		&= \half (\t\p{_i} \t u{_j} + \t\p{_j} \t u{_i})\,,
	&
	\t{\underline \sigma}{^i^j}
	    &= \lambda\, \t \delta{^i^j} \t{\underline e}{^k_k}
	        + 2 \mu \t{\underline e}{^i^j}\,.
\end{align}
Using these quantities, one can then decompose the strain $\t e{_i_j}$ and stress $\t \sigma{^i^j}$ into contributions arising from the displacement (with respect to TT coordinates) and the gravitational wave amplitude as
\begin{align}
	\t e{_i_j}
		&= \t{\underline e}{_i_j} + \half \t h{_i_j}
		+ \order{2}\,,
	&
	\t \sigma{^i^j}
		&= \t{\underline \sigma}{^i^j} + \mu \t h{^i^j}
		+ \order{2}\,.
\end{align}
In terms of $\t{\underline e}{_i_j}$ and $\t{\underline \sigma}{^i^j}$, the equations of motion take the form
\begin{align}
	\varrho \t{\ddot u}{^i} - \t\p{_j} \t{\underline \sigma}{^j^i} &= \order{2}\,,
	&
	(\t{\underline \sigma}{^i^j} \t n{_j} + \mu \t h{^i^j} \t n{_j}) |_{\p U} &= \order{2}\,.
\end{align}
In this picture, the gravitational wave thus acts purely as a traction force $\t t{^i} = - \mu \t h{^i^j} \t n{_j}$ on the boundaries of the material.

\section{Longitudinal Oscillations of Thin Rods}
\label{s:thin rods}

To illustrate the above equations, consider a thin elastic rod of length $L$ (which we take to be aligned with the $x$-axis), which is set into longitudinal oscillations by a normally-incident gravitational wave.

For longitudinally oscillating thin rods, $\sigma \equiv \t\sigma{^x^x}$ is the only non-zero component of the stress tensor \cite[§25]{Landau7}. The stress-strain relationship provided by \cref{eq:stress from strain} then implies
\begin{align}
	\t e{_x_y} = \t e{_x_z} &= 0\,,
	&
	\t e{_y_y} = \t e{_z_z} &= - \nu \t e{_x_x}\,,
\end{align}
where $\nu = \frac{\lambda}{2(\lambda + \mu)}$ is Poisson’s ratio. Hence, the entire strain tensor can be expressed using the single component \add{$\varepsilon \equiv \t e{_x_x}$}, which is related to the stress by \add{$\sigma = Y \varepsilon$}, where $Y = \frac{\mu(3 \lambda + 2\mu)}{\lambda + \mu}$ is Young’s modulus.
One then obtains the well-known wave equation for longitudinal vibrations of rods, together with non-trivial boundary conditions arising from the gravitational wave:
\begin{align}
	\varrho \t{\ddot u}{^x} - Y \p_x^2 \t u{^x} &= 0\,,
	&
	(\t\p{_x}\t u{^x} + \half \t h{_x_x})|_{x = \pm L/2} &= 0\,.
\end{align}
In agreement with the general equations described in the previous section, one finds that (in the coordinate system used) the gravitational wave enters as an effective surface traction, but produces no effective bulk force.

An alternative description is obtained by writing the displacement $u$ in the form
\begin{equation}
	\t u{^x}(t,x) = - \half \t h{_x_x}(t) x + \delta u(t,x)\,.
\end{equation}
Inserting this into the equations above, one obtains
\begin{align}
	\varrho \delta \ddot u - Y \p_x^2 \delta u &= \half \varrho \p_t^2 \t h{_x_x} x\,,
	&
	(\t\p{_x}\delta u)|_{x = \pm L/2} &= 0\,.
\end{align}
These equations are equivalent to those of the classical theory, where no \add{surface traction is present}, but where there is a bulk acceleration of $\half x \p_t^2 \t h{_x_x} = - x \t R{_0_x_0_x}$, where $\t R{_a_b_c_d}$ is the Riemann curvature tensor of space-time.

The physical length of the rod (measured at constant $t$) then takes the form
\begin{equation}
    \begin{split}
    \ell_\text{ph.}
        &= L [1 + \half \t h{_x_x}(t)] + u(+L/2) - u(-L/2) + \order{2}\\
        &= L + \delta u(+L/2) - \delta u(-L/2) + \order{2}\,,
    \end{split}
\end{equation}
which shows that $\delta u$ describes the absolute physical elongation of the rod.

To illustrate these equations, consider the steady-state solution corresponding to a monochromatic gravitational wave with amplitude $h$ (the general case can be obtained by Fourier synthesis) ${\t h{_x_x} = h \cos(\omega t)}$:
\begin{align}
    \t u{^x}
        &= - \quarter h L \frac{\sin(2 \zeta x/L)}{\zeta \cos(\zeta)} \cos(\omega t)\,,
    &
    \ell_\text{ph.}
        &= L [1 + \half h (1- \tan \zeta / \zeta) \cos(\omega t)]\,,
\end{align}
where $\zeta = L \omega / 2 v$ is a dimensionless parameter, in which  $v = \sqrt{Y/\varrho}$ is the speed of sound for longitudinal oscillations.

\section{Comparison with Synge--Bennoun Theory}
\label{s:CQ vs SB}
In Ref.~\cite{AIHPA_1972__16_1_63_0}, Papapetrou considered the same problem of gravitational-wave induced elastic oscillations in three-dimensional materials from the point of view of the SB theory.
This theory does not contain a notion of deformation \emph{per se}, but works instead with a deformation-rate tensor $\t{\dot e}{_a_b} = \half \LieD_v \t\metricL{_a_b}$, where $\mathfrak L_v$ denotes the Lie derivative along the vector field $\t v{^a}$, defined as
\begin{equation}
    \begin{split}
    \LieD_v \t X{^a^b^\cdots_c_d_\cdots}
      = \t v{^e} \t\nabla{_e} \t X{^a^b^\cdots_c_d_\cdots}
      &
      - (\t\nabla{_e} \t v{^a}) \t X{^e^b^\cdots_c_d_\cdots}
      - (\t\nabla{_e} \t v{^b}) \t X{^a^e^\cdots_c_d_\cdots}
      - \cdots
      \\&
      + (\t\nabla{_c} \t v{^e}) \t X{^a^b^\cdots_e_d_\cdots}
      + (\t\nabla{_d} \t v{^e}) \t X{^a^b^\cdots_c_e_\cdots}
      + \cdots\,.
    \end{split}
\end{equation}
It should be noted that Papapetrou originally used different symbols for these quantities; the notation here was adapted to avoid confusion with the one of Carter and Quintana.
Lacking a notion of strain, this theory cannot formulate a stress-strain relationship directly, but rather defines a spatial stress tensor $\t\theta{_a_b}$ via the evolution equation
\begin{equation}
    \LieD_v \t\theta{_a_b} = \t\Gamma{_a_b^c^d} \t{\dot e}{_c_d}\,,
\end{equation}
where $\t\Gamma{_a_b^c^d}$ is a spatial stiffness tensor.

In the theory developed here, however, one has
\begin{equation}
    \mathfrak L_v \t C{^a^b^c^d}
        = \t v{^a} \t a{_{a'}} \t C{^{a'}^b^c^d}
        + \t v{^b} \t a{_{b'}} \t C{^a^{b'}^c^d}
        + \t v{^c} \t a{_{c'}} \t C{^a^b^{c'}^d}
        + \t v{^d} \t a{_{d'}} \t C{^a^b^c^{d'}}\,,
\end{equation}
where $\t a{^b} = \t v{^a} \t\nabla{_a} \t v{^b}$ is the four-acceleration.
Consequently, one has
\begin{equation}
    \mathfrak L_v \t \sigma{^a^b}
        = \t C{^a^b^c^d} \t{\dot e}{_c_d}
        + \t v{^a} \t \sigma{^b^c} \t a{_c}
        + \t v{^b} \t \sigma{^a^c} \t a{_c}\,,
\end{equation}
or equivalently
\begin{equation}
    \mathfrak L_v \t \sigma{_a_b}
        = \t C{_a_b^c^d} \t{\dot e}{_c_d}
        + \t v{_a} \t \sigma{_b_c} \t a{^c}
        + \t v{_b} \t \sigma{_a_c} \t a{^c}
        + \t k{_a_c} \t \sigma{^c_b}
        + \t k{_b_c} \t \sigma{^c_a}
        \,,
\end{equation}
where $\t k{_a_b} = \LieD_v \t g{_a_b} = \t\nabla{_a} \t v{_b} + \t\nabla{_b} \t v{_a}$.

Accordingly, Papapetrou’s stress tensor $\t\theta{^a^b}$ for the SB theory differs from the Cauchy stress tensor $\t\sigma{^a^b}$ derived from the CQ theory.
However, in the present setup, the terms $\t \sigma{^a^b}$, $\t a{^b}$, and $\t k{_a_b}$ are of first order in the expansion parameter $\epsilon$ (measuring the strength of the gravitational wave) so that in this specific case one can set $\t\Gamma{_a_b^c^d} = \t C{_a_b^c^d} + O(\epsilon)$ to obtain
\begin{equation}
    \mathfrak L_v \t\theta{_a_b} = \mathfrak L_v \t\sigma{_a_b} + \order{2}\,.
\end{equation}
If one requires both $\t\theta{_a_b}$ and $\t\sigma{_a_b}$ to vanish before the arrival of the gravitational wave, this implies
\begin{equation}
    \t\theta{_a_b} = \t \sigma{_a_b} + \order{2}\,.
\end{equation}
Hence, in the linearised scheme, both stress tensors coincide (to the relevant order).
So while Papapetrou’s theory is based on different methods and assumptions, its predictions coincide with those derived from the elasticity theory of Carter and Quintana.

\section{Discussion}

The theory of relativistic elasticity thus provides a rigorous derivation of the previously postulated elasticity equations used by previous authors \cite[Sect.~8.1]{Maggiore2008}.

As the current theories of relativistic elasticity are limited to hyperelasticity, so is the model considered here.
More physically \add{plausible} models would have to contain descriptions of thermo-elastic and visco-elastic effects to describe dissipation and friction.
While such terms can be added to the equations given here, following the non-relativistic theory, we regard search \add{modifications} as ad-hoc.
Instead, it would be of interest to extend the current theory of relativistic elasticity to allow for materials which are not hyperelastic, but also have thermo-elastic and visco-elastic properties.
A logically consistent extension of the equations given here should then be obtainable from this future elasticity theory via a perturbative expansion similar to the one used here.

Independent of the generalisation to more accurate matter models, it would be of interest to extend the analysis presented here beyond the one-dimensional case, in order to realistically describe the response of two-dimensional plates or three-dimensional objects to gravitational radiation. This could serve as a basis for rigorous theoretical models of Weber-bar-type gravitational wave detectors and the proposed \emph{Lunar Gravitational-Wave Antenna} \cite{2021ApJ...910....1H}. Conceptually, it could also be of interest to consider astrophysical applications of this theory, e.g.\ the excitation of elastic oscillations of neutron stars in the vicinity of binary black-hole mergers, due to gravitational radiation. However, as the computation of gravitational wave effects on two-dimensional systems is already quite demanding \cite{Spanner_2022}, it seems probable that such models will not be analytically solvable, but require numerical methods instead.

Further, it  would certainly be of relevance to extend our analysis to pre-stressed materials, and to study the above-mentioned scenarios in the full non-linear elasticity theory as well.

\section*{Acknowledgements}

We thank Piotr Chruściel and Robert Beig for helpful discussions.
T.M. is a recipient of a DOC Fellowship of the Austrian Academy of Sciences at the University of Vienna, Faculty of Physics and is supported by the Vienna Doctoral School in Physics (VDSP).

\printbibliography[heading=bibintoc]
\end{document}